\documentstyle[aps,prl,floats,amssymb,array]{revtex}
\def\z{{\Bbb Z}}

\def\abs#1{ \left| #1 \right | }
\def\vev#1{ \langle #1 \rangle  }

\newcommand{\drawsquare}[2]{\hbox{%
\rule{#2pt}{#1pt}\hskip-#2pt
\rule{#1pt}{#2pt}\hskip-#1pt
\rule[#1pt]{#1pt}{#2pt}}\rule[#1pt]{#2pt}{#2pt}\hskip-#2pt
\rule{#2pt}{#1pt}}

\newcommand{\Yfund}{\raisebox{-.5pt}{\drawsquare{6.5}{0.4}}}
\newcommand{\Ysymm}{\raisebox{-.5pt}{\drawsquare{6.5}{0.4}}\hskip-0.4pt%
        \raisebox{-.5pt}{\drawsquare{6.5}{0.4}}}
\newcommand{\Ythrees}{\raisebox{-.5pt}{\drawsquare{6.5}{0.4}}\hskip-0.4pt%
          \raisebox{-.5pt}{\drawsquare{6.5}{0.4}}\hskip-0.4pt%
          \raisebox{-.5pt}{\drawsquare{6.5}{0.4}}}
\newcommand{\Yasymm}{\raisebox{-3.5pt}{\drawsquare{6.5}{0.4}}\hskip-6.9pt%
        \raisebox{3pt}{\drawsquare{6.5}{0.4}}}
\newcommand{\Ythreea}{\raisebox{-3.5pt}{\drawsquare{6.5}{0.4}}\hskip-6.9pt%
        \raisebox{3pt}{\drawsquare{6.5}{0.4}}\hskip-6.9pt
        \raisebox{9.5pt}{\drawsquare{6.5}{0.4}}}
\newcommand{\Yfoura}{\raisebox{-3.5pt}{\drawsquare{6.5}{0.4}}\hskip-6.9pt%
        \raisebox{3pt}{\drawsquare{6.5}{0.4}}\hskip-6.9pt
        \raisebox{9.5pt}{\drawsquare{6.5}{0.4}}\hskip-6.9pt
        \raisebox{16pt}{\drawsquare{6.5}{0.4}}}
\begin{document}

\tightenlines

\twocolumn[\hsize\textwidth\columnwidth\hsize\csname@twocolumnfalse\endcsname

\preprint{\vbox{\hbox{UCSD/PTH 97--35}}}

\title{Supersymmetric Gauge Theories with an Affine Quantum Moduli Space}
\author{Gustavo Dotti and Aneesh V.~Manohar}
\address{Department of Physics, University of California at San Diego,\\
9500 Gilman Drive, La Jolla, CA 92093-0319}
\date{November 1997}
\maketitle

\begin{abstract} 
All supersymmetric gauge theories based on simple groups which have an affine
quantum moduli space, i.e.\ one generated by gauge invariants with no
relations, $W=0$, and anomaly matching at the origin, are classified.  It is
shown that the only theories with no gauge invariants (and moduli space equal
to a single point) are the two known examples, $SU(5)$ with $\overline 5 + 10$
and $SO(10)$ with a spinor. The index of the matter representation must be at
least as big as the index of the adjoint in theories which have a non-trivial
relation among the gauge invariants.
\end{abstract}
\pacs{PACS}

]\narrowtext

The moduli space of supersymmetric gauge theories is described in terms of
gauge invariant composite fields made out of the microscopic fields.  In
general, the low energy theory can have a superpotential constructed from the
composite fields, as well as non-trivial polynomial relations among the
composite fields. The structure of some of these moduli spaces has been studied
in detail~\cite{seiberg,susy}. In this paper, we classify all supersymmetric
gauge theories based on simple groups with an affine quantum moduli space.
These are theories in which the moduli space is given by gauge invariant
polynomials with no relations between them, and has $W=0$. The flavor anomalies
of the fundamental fields agree with those computed using the gauge invariant
composites at all points, including the origin. The low-energy massless modes
are given by the gauge invariant composites, and are the same at every point on
the moduli space. Some of these theories have several branches, of which at
least one has $W=0$. We find two side results: (i) the only theories with
simple gauge groups for which there are no gauge invariant composite fields are
the two cases known in the literature, $SU(5)$ with $\bar 5 + 10$, and $SO(10)$
with a single spinor~\cite{su5so10}.  The moduli space of these theories is a
single point, and the theories are expected to break supersymmetry dynamically.
(ii) Theories with non-trivial relations among the basic gauge invariants must
have $\mu \ge \mu_{\rm adj}$.  (Here $\mu$ is the index of the matter
representation, and $\mu_{\rm adj}$ is the index of the adjoint.)

The most familiar example of a supersymmetric gauge theory is supersymmetric
QCD with $N_C$ colors and $N_F$ flavors of  quarks $Q^{i\alpha}$ and antiquarks
$\tilde Q_{j\beta}$. This theory does not satisfy the criteria for an affine
moduli space: it either has a superpotential, a relation between the
composites, or both. The classical  moduli space for $N_F < N_C$ is given by
the value of the meson field $M^i_j= Q^{i\alpha}\tilde Q_{j\alpha}$, and the
quantum theory has a dynamically generated superpotential~\cite{seiberg}
\begin{equation}\label{1}
W=(N_C-N_F) \left[ {\Lambda^{3N_C - N_F} \over \det M} \right]^{1/ \left (
N_C-N_F \right)}.
\end{equation}
When $N_F=N_C$, the moduli space is given by the values of the meson $M$, and
baryons $B = \det Q$ and $\tilde B = \det \tilde Q$, subject to the quantum
constraint $\det M - B \tilde B = \Lambda^{2 N_C}$, and $W=0$. For $N_F=N_C+1$,
the moduli space is given by the superpotential~\cite{seiberg}
\begin{equation}\label{2}
W={ \tilde B_i M^i_j B^j - \det M \over \Lambda^{2 N_C -1 }}.
\end{equation}
The gauge invariants, $M^i_j$, $B^j$ and $\tilde B_i$ have relations among them
(such as $M^i_j B^j=0$) which are obtained by varying the superpotential $W$ in
Eq.~(\ref{2}). If $N_F > N_C+1$, the low energy theory has a dual
description~\cite{seiberg}.  Another interesting family of theories is 
supersymmetric $SO(N)$ gauge theory with $N-4$ flavors of matter in the vector
representation $Q_{i \alpha}$~\cite{SOn}.  In this case, the moduli space is
described by the value of the meson field $M_{(ij)}=Q_{i \alpha} Q_{j\alpha}$.
There are four branches, two with $W=0$, and the other two with $W = \pm {2
\Lambda^{N-1}/\sqrt{ \det M } }$. The $W=0$ branch is an example of an affine
moduli space.

The classification of all supersymmetric gauge theories with an affine quantum 
moduli space is straightforward, but tedious.  All maximal representations
$\rho_{\rm max}$ of simple Lie groups $G$ with a free algebra of $G$-invariant
polynomials have already been classified in the mathematics
literature~\cite{schwarz,adamovich}. (A free algebra is one in which there are
no relations among the generators.) It remains to look at all subsets of
$\rho_{\rm max}$, and to check that the theory has no gauge anomalies and is
asymptotically free, and that the flavor anomalies satisfy 't~Hooft's
consistency conditions~\cite{thooft}. In particular, one requires that the
anomalies match at the origin. The quantum theory is then expected to be a
confining theory, and the low-energy dynamics is given by a supersymmetric
effective Lagrangian written in terms of the composite fields, with a Kahler
potential that is smooth at the origin. One can also classify all theories
which have no gauge invariant composites, since these are a special case of
theories with a free algebra of invariants. For these theories, one does not
have any anomaly matching constraints.  The resulting theories (after checking
$\sim 200$ cases) are listed in Table~\ref{tab:1}, and the invariants are
listed in Table~\ref{tab:2}. The theories can be divided into three groups:
T1--T6 have $\mu < \mu_{\rm adj}$, T7--T11 have $\mu > \mu_{\rm adj}$. Theories
S1 and S2 have no invariant polynomials. These have been studied
before~\cite{su5so10}, and are believed to break supersymmetry dynamically.
These are the only two theories based on simple groups which have no gauge
invariants, and whose moduli space is a single point. One can also check that
all physically interesting theories, i.e. those with no gauge anomalies, with
$\mu < \mu_{\rm adj}$ ($\mu \le \mu_{\rm adj}$ if the representation is
irreducible) have a free algebra of invariants.  Thus all theories with
constraints must have $\mu \ge \mu_{\rm adj}$. This is true even if the
relation involves the non-perturbative scale $\Lambda$, because for large
values of the fields, $\Lambda$ can be neglected, and the quantum relation
reduces to a classical relation.

Theories with $\mu < \mu_{\rm adj}$ can have a dynamically generated
superpotential. The general form of a superpotential consistent with the R
symmetry is a sum of terms of the form
\begin{equation}
W = \Lambda^{\left( \mu - 3 \mu_{\rm adj} \right)/ \left( \mu - \mu_{\rm adj} \right)}
\Pi_i\ \phi_i^{2 \mu_i/\left( \mu - \mu_{\rm adj} \right)},
\end{equation}
where $\phi_i$ are the elementary fields with index $\mu_i$. The product of
fields $\phi_i$ must be gauge and flavor invariant. For asymptotically free theories, $\mu
< 3 \mu_{\rm adj}$, so that the power of $\Lambda$ is positive if $\mu <
\mu_{\rm adj}$. This means that far away from the origin of moduli space, where
the classical description is valid, the superpotential $W \rightarrow 0$, as
one would expect in the classical theory. For $\mu > \mu_{\rm adj}$, $\Lambda$
occurs with a negative power. This is not an acceptable form for $W$, since $W
\rightarrow \infty$ for large values of the fields, which disagrees with the
classical result. There is a loophole to the above argument, since
supersymmetric QCD with $N_F=N_C+1$ has $W$ of the form Eq.~(\ref{2}), with
$\Lambda$ in the denominator. In this case, the numerator of $W$ vanishes on
the moduli space, because of the constraint equations between the mesons and
baryons, so that $W \rightarrow 0$ at infinity. However, the theories we are
considering are precisely those which have no relations among the gauge
invariants, so one cannot have a negative power of $\Lambda$ in $W$, and there
is no dynamically generated superpotential if $\mu > \mu_{\rm adj}$.

The first six theories all have $\mu < \mu_{\rm adj}$ and can have a
dynamically generated superpotential. They all have several branches, with a
branch having $W=0$ and an affine moduli space. This has been well-studied in
the case of $SO(N)$ gauge theory with $N_F=N-4$ flavors of matter $Q_{ia}$ in
the fundamental representation~\cite{SOn}. The moduli space is described by the
meson fields $M_{ij} = Q_{i\alpha} Q_{j\alpha}$, which is an $N_F \times N_F$
symmetric matrix. The 't~Hooft conditions are satisfied at all points of the
moduli space, including the origin. There are points on the moduli space where
the gauge group is spontantaneously broken to $SO(4)\sim SU(2) \times SU(2)$ by
giving vacuum expectation values to the matter fields. Each $SU(2)$ gauge
theory has superpotential $W = \pm \Lambda_L^3$, where the two possible signs
correspond to the two different vacua of the $SU(2)$ gauge theory, and
$\Lambda_L$ is the low-energy scale parameter of the $SU(2)$ theory. The total
$W$ has the form $W = \pm \Lambda_L^3 \pm \Lambda_L^3$ from the two $SU(2)$'s,
which have identical couplings, and hence identical $\Lambda_L$'s.  Matching
the gauge couplings of the original $SO(N)$ theory to the $SO(4)$ coupling
gives the relation $\Lambda_L^3 = \Lambda^{N-1}/ \sqrt {\det M}$, where
$\Lambda$ is the scale parameter of the $SO(N)$ theory. Thus the superpotential
of the $SO(N)$ theory has the values $W=0, \pm 2\Lambda^{N-1} /\sqrt {\det M}$.
One can add a mass term $W_m = m_{ij} Q_{i\alpha} Q_{j\alpha} = m_{ij} M_{ij}$
to the original theory. The branch with $W=0$ now has $W=m_{ij} M_{ij}$,  and
so has no supersymmetric ground state if one assumes that the Kahler potential
is smooth at the origin when written in terms of $M$. This does not mean that
supersymmetry is broken: there is a non-trivial solution to $\partial
W/\partial M=0$ in the branches $W = \pm 2\Lambda^{N-1}/\sqrt {\det M} + m_{ij}
M_{ij}$. In summary, $SO(N)$ with $(N-4) \Yfund$ has multiple branches, two of
which have $W=0$. Perturbing the microscopic theory by a mass term  lifts the
$W=0$ branches, but there  are still supersymmetric solutions from the
$W\not=0$ branches of the theory.

One expects the other five theories with $\mu < \mu_{\rm adj}$ to also be
multi-branched theories, with a branch having $W=0$. There are several ways to
check that this is true. One way is to note that the remaining five theories
can all be obtained from s-confining theories~\cite{sconf}  by integrating out
matter. It is straightforward to  verify that one gets several branches, one of
which has $W=0$. The same result can also be obtained by looking at gaugino
condensation. In T6, $SO(14) \rightarrow G_2 \times G_2$, one can get $W=0$
from a cancellation between the superpotentials $W=\omega_4^r \Lambda^3$
($\omega_4$ is a fourth root of unity) due to gaugino condensation in each
$G_2$. The origin of $W=0$ for the other theories is more subtle. In T2, one
gets two unbroken $SU(3)$ gauge theories. Each $SU(3)$ gauge theory has
$W=\Lambda_L^3 \omega^r$, where $\omega$ is a cube root of unity. Naively, one
expects that the superpotential is the sum $W = \Lambda_L^3 \left(\omega^r +
\omega^s \right)$ of the two $SU(3)$'s, which does not have a $W=0$ branch.
However, a more careful analysis shows that $W$ is the difference  $W =
\Lambda_L^3 \left(\omega^r - \omega^s \right)$, which does have a $W=0$ branch.
One can see this by studying the breaking of $SU(6) \rightarrow SU(3) \times
SU(3)$. The expectation value $\vev{A_{\left[123\right]}}=v_1$, $\vev{A_{\left[
456 \right]}}=v_2$ breaks $SU(6) \rightarrow SU(3) \times SU(3)$, and one needs
$\abs{v_1} = \abs{v_2}$ to satisfy the $D$-flatness condition. The matching
condition on $\Lambda$ is $\Lambda_{Li}^3 = \Lambda^5/v_i^2$. One can
interchange the two $SU(3)$ groups by acting with the $SU(6)$ matrix
\begin{equation}
U=\left(\begin{array}{cccccc}
0 & 0 & 0 & i & 0 & 0\\
0 & 0 & 0 & 0 & i & 0\\
0 & 0 & 0 & 0 & 0 & i\\
i & 0 & 0 & 0 & 0 & 0\\
0 & i & 0 & 0 & 0 & 0\\
0 & 0 & i & 0 & 0 & 0\\
\end{array}\right),
\end{equation}
which maps $v_1 \rightarrow -i v_2$, $v_2 \rightarrow -i v_1$. The factors of $i$
are necessary for the matrix to have $\det U = 1$. Under $U$, one finds that
$\Lambda_{L 1,2}^3 \rightarrow - \Lambda_{L 2,1}^3$, so that it is the
difference of $W$'s which is $SU(6)$ invariant. 

The origin of $W=0$ for T1 is also interesting, especially when $N$ is odd,
since then one has the sum of an odd number of $SU(2)$ superpotentials. This
can be analyzed for the case of $SU(6) \rightarrow \left( SU(2) \right)^3$ by
the vacuum expectation value of $A$ and $A^*$, the two matter fields in the
antisymmetric representation. The breaking is due to
\begin{equation}
\vev{A}=
\left(\begin{array}{cccccc}
0 & v_1 & 0 & 0 & 0 & 0\\
-v_1 & 0 & 0 & 0 & 0 & 0\\
0 & 0 & 0 & v_2 & 0 & 0\\
0 & 0 & -v_2 & 0 & 0 & 0\\
0 & 0 & 0 & 0 & 0 & v_3\\
0 & 0 & 0 & 0 & -v_3 & 0\\
\end{array}\right),\
\end{equation}
\begin{equation}
\vev{\tilde A}=\left(\begin{array}{cccccc}
0 & \tilde v_1 & 0 & 0 & 0 & 0\\
-\tilde v_1 & 0 & 0 & 0 & 0 & 0\\
0 & 0 & 0 & \tilde v_2 & 0 & 0\\
0 & 0 & -\tilde v_2 & 0 & 0 & 0\\
0 & 0 & 0 & 0 & 0 & \tilde v_3\\
0 & 0 & 0 & 0 & -\tilde v_3 & 0\\
\end{array}\right),\
\end{equation}
with $v_i^2 - \tilde v_i^2 = {\rm constant}$, which breaks $SU(6) \rightarrow
\left( SU(2) \right)^3$ as long as the $v_i^2$ are all different. The masses of
the gauge bosons corresponding to the broken generators are proportional to the
differences $v_i^2-v_j^2,\ i \not=j$. The relations between the low-energy
$\Lambda_i$'s and $\Lambda$, obtained by matching coupling constants at the
gauge boson mass scales, are: $\Lambda_1^3 = {\Lambda^7 /\left( v_1^2-v_2^2
\right) \left(v_1^2 - v_3^2 \right) }$, $\Lambda_2^3 = {\Lambda^7 /\left(v_2^2
- v_1^2 \right) \left(v_2^2 - v_3^2 \right)}$, $\Lambda_3^2 = {\Lambda^7 /
\left( v_3^2-v_1^2 \right)  \left(v_3^2 - v_2^2 \right) }$. The $SU(6)$
superpotential is the sum of the three $SU(2)$ superpotentials,
\begin{eqnarray}
W &=& \pm {\Lambda^7 \over \left( v_1^2-v_2^2 \right) \left(v_1^2 - v_3^2
\right) } \pm {\Lambda^7 \over \left(v_2^2 - v_1^2 \right)
\left(v_2^2 - v_3^2 \right)} \nonumber\\
&&\qquad \pm {\Lambda^7 \over \left( v_3^2-v_1^2 \right) 
\left(v_3^2 - v_2^2 \right) },
\end{eqnarray}
and has a $W=0$ branch~\cite{skiba}. The other $SU(2N)$ T1 theories can be
analyzed similarly. An almost identical analysis also explains the $W=0$ branch
for the $Sp(2N)$ theory T3. This theory has been analyzed previously in
Ref.~\cite{cho}, where the $W=0$ branch was obtained by integrating out matter.

Theories T7--T11 have $\mu > \mu_{\rm adj}$, and so cannot have a dynamically
generated superpotential. Theory T7 has been studied before~\cite{iss}.  It has
no quadratic invariant, and so is a chiral theory. It is expected to break
supersymmetry dynamically when the microscopic theory is perturbed by adding $m
\phi^4=m u$ to $W$.

Theories T8--T11 all have quadratic invariants and are not chiral theories,
since one can give mass to all the microscopic matter fields. One can study the
low-energy behavior of these theories by studying the flows at certain points
on the moduli space.  Theory T11 flows to T8 and T9 flows to $SO(6)$ with
$\Ysymm$ when the matter fields get vacuum expectation values. Thus it is
sufficient to understand the low energy behavior of theories T8 and T10.

The $SU(8)$ theory with the four-index antisymmetric tensor flows, via the
Higgs mechanism, to $SU(4) \times SU(4)$ with $(\Yasymm,\Yasymm)$, which is the
same as $SO(6) \times SO(6)$ with $(\Yfund,\Yfund)$. One can study this model
in the limit that the two $SO(6)$ couplings $g_1$ and $g_2$ are very different,
$g_1 \gg g_2$. In this case, the strongly coupled $SO(6)$ has 6 flavors of
matter $q$ in the vector representation. This is dual to an $SO(4)$ theory with
6 flavors of vector matter $\tilde q$, gauge singlet fields $M$ that are a
symmetric tensor under flavor, and a superpotential $W=M\tilde q \tilde
q$~\cite{SOn}. At a generic point on the moduli space, the $\tilde q$ fields
are heavy, and one is left with the gauge singlet fields $M$. Including the
dynamics of the weakly coupled $SO(6)$, one finds that one has an $SO(6)$
theory with matter $M$ in the symmetric tensor representation. Thus
understanding theories T8--T11 reduces to understanding the dynamics of $SO(N)$
gauge theories with matter in the symmetric tensor representation.

The $SO(N)$ theory with one $\Ysymm$ is a theory whose low-energy behavior is
not understood. One can perturb the original theory by adding a mass term for
the symmetric tensor, $W=m \phi_{ab} \phi_{ab}$. This gives a superpotential in
the low energy theory, $W=m u_2$, where $u_2=\phi_{ab} \phi_{ab}$ is one of the
basic gauge invariants. If one assumes that the Kahler potential is smooth at
the origin, then supersymmetry must be dynamically broken, since there is no
solution to $\partial W/\partial M = m =0$ for non-zero mass $m$. This looks
very similar to the $SU(2)$ theory with one $\Ythrees$. There are however, some
important differences: one can add a gauge invariant mass term, so the $SO(N)$
theory is not chiral. Also, by the Higgs mechanism, one flows from $SO(N)$ with
$\Ysymm$ to $SO(N-1)$ with $\Ysymm$, and so on down to $SO(4)$ with $\Ysymm$
(or $SO(3)$ with $\Ysymm$) which is not asymptotically free, so the low energy
theory has free quarks and gluons and cannot break supersymmetry
dynamically~\cite{bic}.

In summary, we have found all supersymmetric gauge theories based on simple
groups which have an affine quantum moduli space. The low energy dynamics is
described, except for those theories which reduce to $SO(N)$ with a symmetric
tensor. The dynamics of the $SO(N)$ with $\Ysymm$ theory, as well as the gauge
theories whose chiral ring is a free algebra but do not satisfy the 't~Hooft
consistency conditions, is being studied.

We are indebted to  K.~Intriligator and W.~Skiba for numerous discussions. This
work was supported in part by a Department of Energy grant  DOE-FG03-97ER40546.

\onecolumn
\widetext

\begin{table}[tbh]
\caption{\label{tab:1}
\noindent All theories with unconstrained moduli spaces and $W=0$.  The
anomalies computed using the microscopic fields $\rho$, and using the gauge
invariant composites listed in Table~\ref{tab:2} agree at all points, including
the origin for T1--T11. Theories S1 and S2 have no gauge invariants, and are
the only two theories of this type. Anomaly matching does not hold for these
theories. We omit theories which are conjugate to those listed, or with $S
\rightarrow S'$. Notation: $G$ is the gauge group, $\rho$ the matter
representation ($S$ denotes the spinor representation), $d_\rho$ the dimension
of $\rho$, $d_M$ the number of gauge invariant composites, and $G_*$, the
unbroken gauge group at $D$-flat points where $G$ is maximally broken.  In T10,
$k=N-2$ if $N$ is even, and $k=N-1$ if $N$ is odd. Theory T4 with $N=6$ is
equivalent to T1 with $N=2$. T1--T6 have $\mu < \mu_{\rm adj}$, T7--T11 have 
$\mu > \mu_{\rm adj}$. T10 with $N=3$ satisfies all the anomaly matching
conditions, but is not an asymptotically free theory.}
\setlength{\tabcolsep}{1mm}
\renewcommand{\arraystretch}{1.5}
\newcolumntype{C}{>{$}c<{$}}
\begin{tabular}{|C|C|C|C|C|C|C|C|} \hline 
&G & \rho & d_\rho & d_M & \mu & \mu_{\rm adj} & G_*\\ \hline \hline
T1&SU(2N) & \Yasymm + \overline{\Yasymm} & 2N(2N-1) & N+1
& 2N-2 & 2N & \left( SU(2) \right)^{N}\\
T2&SU(6) & \Ythreea & 20 & 1 & 3 & 6 & SU(3) \times SU(3)\\
T3&Sp(2N), N \geq 2 & \Yasymm & (N-1)(2N+1)& N-1 & N-1 & N+1 &
\left( SU(2) \right)^{N}\\ 
T4&SO(N), N \geq 5 & (N-4)\Yfund & N(N-4)& \frac{1}{2}(N-4)(N-3) &
N-4 & N-2 & SU(2) \times SU(2)\\
T5& SO(12) & 2 S &  64 & 7 & 8 & 10 &\left( SU(2) \right)^3\\
T6&SO(14)& S & 64 & 1 & 8 & 12 & G_2 \times G_2\\
\hline
T7&SU(2) & \Ythrees & 4 & 1 & 5 & 2 &\z_3\\
T8&SU(8) & \Yfoura & 70 & 7 & 10 & 8 & (\z_2)^6\\
T9&Sp(8) & \Yfoura & 42 & 6 & 7 & 5 & (\z_2)^6\\
T10&SO(N), N \geq 5 & \Ysymm & 
 \frac{1}{2}(N-1)(N+2)& N-1 & N+2 & N-2 & (\z_2)^k\\
T11&SO(16)& S & 128 & 8 & 16 & 14 &(\z_2)^8 \\ 
\hline\hline
S1&SU(5) & \overline{\Yfund} + \Yasymm & 15 & 0 & 2 & 5 & SU(5)\\
S2&SO(10) & S & 16 & 0 & 2 & 8 & SO(10)\\
\hline
\end{tabular}
\end{table}

\begin{table}[tbh]
\caption{\label{tab:2} Invariants for the theories in Table~\ref{tab:1}.
Details about the index contractions have been omitted. Flavor indices are
denoted by $i$, and gauge indices by Greek letters.}
\setlength{\tabcolsep}{3mm}
\renewcommand{\arraystretch}{1.25} 
\newcolumntype{C}{>{$}c<{$}}
\newcolumntype{L}{>{$}l<{$}}
\begin{tabular}{|C|L|L|}\hline
& {\rm Fields} & {\rm Invariants} \\
\hline\hline
T1& A_{[\alpha\beta]},\ \tilde A^{[\alpha\beta]}& u_k=\left(A \tilde A
\right)^k,\ k=1,\ldots,N-1;\ b = {\rm Pf} A;\ \tilde b = {\rm Pf} \tilde A \\
T2& A_{[\alpha\beta\gamma]} & u = A^4 \\
T3& A_{[\alpha\beta]} & u_k = A^k,\ k =2,\ldots,N \\ 
T4& Q_{i\alpha} & M_{ij} = Q_{ia}Q_{ja}  \\
T5& \phi_{i\alpha} & u_k=\left( \phi_1 \phi_2 \right)^k,\ k=1,3;\ 
w_r = \phi_1^r \phi_2^{4-r}, 0 \le r \le 4  \\
T6& \phi_{\alpha} & u=\phi^8 \\
\hline
T7& \phi_{(\alpha\beta\gamma)} & u = \phi^4  \\
T8& A_{[\alpha\beta\gamma\lambda]} & u_k = A^k,\ k=2,6,8,10,12,14,18  \\
T9& A_{[\alpha\beta\gamma\lambda]} & u_k = A^k,\ k=2,5,6,8,9,12  \\
T10& \phi_{\alpha\beta} & u_k = \phi^k,\ k=2,3,\ldots,N  \\
T11& \phi_\alpha & u_k = \phi^k,\ k=2,8,12,14,18,20,24,30  \\ 
\hline
\end{tabular}
\end{table}


\begin{references}
\bibitem{skiba} This example was worked out with the help of W.~Skiba.
\end{references}
\end{document}